\documentclass[12pt]{article}
\usepackage[dvipsnames]{xcolor}
\usepackage{amsmath,amssymb,hyperref,tikz}
\usepackage[utf8]{inputenc}
\newcommand{\beq}{\begin{equation}\begin{aligned}}
\newcommand{\eeq}{\end{aligned}\end{equation}}

\newcommand{\dd}{\mathcal{D}}
\usepackage[top=1in, bottom=1in, left=1.in, right=1.in]{geometry}
\usepackage[english]{babel}
\usepackage{atbegshi,cite}
\usepackage{amsmath,amssymb,amsbsy,amstext, amsthm, simplewick}
\usepackage{hyperref}
\usepackage{graphicx}
\usepackage{amsfonts}
\usepackage[small]{caption}
\usepackage{upgreek}
\usepackage[titletoc]{appendix}
\usepackage{setspace}
\usepackage{slashed}

\newcommand{\newc}{\newcommand}
\newc{\be}{\begin{equation}}
\newc{\ee}{\end{equation}}
\newc{\fpi}{f_{\pi}}
\newc{\etap}{\eta^{\prime}}
\newc{\llll}{\langle\lambda\lambda\rangle}
\newc{\FFd}{F^a\tilde F^a}
\newc{\qbar}{{\overline q}}
\newc{\TR}{{\rm Tr}}
\newc{\Kahler}{K\"ahler }
\newc{\Zbb}{{\mathbb Z}}
\newc{\Rt}{{\mathbb R}^3}
\newc{\Rf}{{\mathbb R}^4}
\newc{\So}{{\mathbb S}^1}
\newc{\zt}{{\mathbb Z}_2}
\newc{\RtSo}{{\mathbb R}^3\times{\mathbb S}^1}
\newc{\scriminus}{{\cal I}^-}
\newc{\scriplus}{{\cal I}^+}
\newc{\mpl}{M_p}
\newc{\Ricci}{\mathcal{R}}
\newc{\bv}{\phi}
\newc{\calU}{{\cal U}}
\newc{\calK}{K}
\newc{\calUi}{{\cal U}^{-1}}
\newc{\calG}{{\cal G}}
\newc{\calM}{{\cal M}}
\newc{\calL}{{\cal L}}
\newc{\calO}{{\cal O}}
\newc{\calR}{{\cal R}}
\newc{\calQ}{{\cal Q}}
\newc{\calI}{{\cal I}}
\newc{\calOb}{{\cal O}^\dagger}
\newc{\hphi}{{\hat\phi}}

\newc{\tb}[1]{{\bf\color{MidnightBlue}TB:~#1}}
\newc{\pd}[1]{{\bf\color{blue}PD:~#1}}

\theoremstyle{plain}
\theoremstyle{plain} 
\theoremstyle{plain} 
\theoremstyle{plain}
\theoremstyle{plain}
\theoremstyle{plain}

\renewcommand{\title}[1]{{\Large\bf\flushleft{#1}}\vspace*{3ex}\\}
\renewcommand{\author}[2]{{\noindent\hspace*{2.5em}\large#1}
                     \footnote{Electronic mail: $\mathtt{#2}$}\\}

\begin{document}
\begin{titlepage}
\begin{flushright}
{\large 
~\\
}
\end{flushright}

\vskip 2.2cm

\begin{center}

{\large \bf Lattice BF Theory, Dumbbells, and Composite Fermions}

\vskip 1.4cm

{{Tom Banks and Bingnan Zhang\footnote{bingnan.zhang@rutgers.edu}}}
\\
\vskip 1cm
{Department of Physics and NHETC, \\Rutgers University, Piscataway, NJ 08854}\\
\vspace{0.3cm}
\vskip 4pt

\vskip 1.5cm

\begin{abstract}
We formulate $U(1)$ $bda$ Chern-Simons theory, which is also called BF theory, on a lattice, adapting a method proposed by Kantor and Susskind\cite{ks} for the groups $\mathbb{R}$ and $\mathbb{Z}_N$.  Our method applies to any finite or infinite abelian group. We study the discrete symmetries and use the model to provide a rigorous treatment of the composite fermion theory of the fractional quantum Hall effect (FQHE)\cite{jain}\cite{HLR}\cite{fradkin}, with no ambiguities relating to intersecting Wilson/'t Hooft lines.  We derive Jain's fractions, and one can also calculate corrections to the mean field solution within this framework. We also generalize the formalism to higher form gauge models in arbitrary dimension, and suggest a possible non-Abelian extension.\end{abstract}

\end{center}

\vskip 1.0 cm

\end{titlepage}
\setcounter{footnote}{0} 
\setcounter{page}{1}
\setcounter{section}{0} \setcounter{subsection}{0}
\setcounter{subsubsection}{0}
\setcounter{figure}{0}



\tableofcontents
\section{Introduction}

\ \ \ The problem of formulating $U(1)$ Chern-Simons theory on a lattice has a long and twisted history\cite{1,2,3,4,5,6,7,8,9,10} and there is no satisfactory formulation at present.  By contrast, models with action
\beq I = \int b da , \eeq with two different CS gauge potentials have been latticized\cite{kapustin}.  In exploring the literature on this subject we came across an obscure paper by Kantor and Susskind\cite{ks}, which claimed that the lattice $\int b da$ theory coupled to "dumbbell" particles, which lived on both the lattice and the dual lattice, converged in the continuum limit to $\int a d a$ theory coupled to charges.  We will verify the results of\cite{ks} and extend them to the case of gauge group $U(1)$ .  The groups $\mathbb{R}$ and $\mathbb{Z}_N$ studied in\cite{ks} are both self dual.   The key new feature for $U(1)$ is that, since it is not finite, the dual group whose elements are its irreducible representations is $\mathbb{Z}$.  Therefore, we will take the dual lattice link field $b$ to be a $\mathbb{Z}$ gauge potential.   The Wilson loops of this field are thus parametrized by an angle $\theta \in [0,2\pi]$.   

Calculations in a dumbbell CS theory have no problems with self intersecting loops, because the $a$ loops on the lattice and the $b$ loops on the dual lattice never intersect.  We obtain an unambiguous formula for the expectation values of arbitrary products of $a$ and $b$ loops.  Using this formula we are able to find a rigorous second quantized version of the "dualities" of\cite{jain}\cite{HLR}\cite{fradkin}, relating free fermions to dumbbell fermions in certain representations of the groups $(\mathbb{Z},U(1))$ .  These can then be used to construct Jain's composite fermion mean field theory of the fractional quantum Hall effect.  In principle, our models could enable one to do rigorous calculations of the perturbation series around mean field theory, to verify that corrections are small.  

We also generalize our considerations to higher form gauge theories in higher dimensions, and conclude with speculations about the appropriate generalization to non-Abelian CS models.

\begin{figure}[h]
\centering
\scalebox{1.5}{
\begin{tikzpicture}
\foreach \i in {1,-1}
 \foreach \j in {1,-1}
   \foreach \k in {1,-1}
     {
     \coordinate (A_{\i\j\k}) at (\i,\j,\k);
     }
\draw[fill=green!20,opacity=0.2](A_{1-11})--(A_{1-1-1})--(A_{-1-1-1})--(A_{-1-11})--cycle;
\draw[fill=green!20,opacity=0.2](A_{11-1})--(A_{1-1-1})--(A_{-1-1-1})--(A_{-11-1})--cycle;
\draw[fill=green!20,opacity=0.2](A_{-111})--(A_{-11-1})--(A_{-1-1-1})--(A_{-1-11})--cycle;
\draw[fill=green!20,opacity=0.5](A_{111})--(A_{11-1})--(A_{-11-1})--(A_{-111})--cycle;
\draw[fill=green!20,opacity=0.5](A_{111})--(A_{11-1})--(A_{1-1-1})--(A_{1-11})--cycle;
\draw[fill=green!20,opacity=0.5](A_{-111})--(A_{111})--(A_{1-11})--(A_{-1-11})--cycle;
\draw[line width=0.5mm] (0,0,0)--(-1,1,-1);
\fill(0,0,0)circle(1mm);
\draw[line width=0.5mm,dashed](0,0,0)--(1,1,-1);
\fill(-1,1,-1)circle(1mm);
\fill(1,1,-1)circle(1mm);
\draw[->,line width=0.3mm,dotted](-1,1.5,-1) ..controls (0,2,-1)..(1,1.5,-1);
\node at (0,0,0)[left] {$x$};
\node at (-1,1,-1)[above]{$x_d$};
\node at (1,1,-1)[above]{$x_d+l_d$};

\draw[fill=blue!20,opacity=0.2](2,-2,2)--(2,-2,0)--(0,-2,0)--(0,-2,2)--cycle;
\draw[fill=blue!20,opacity=0.2](2,0,0)--(2,-2,0)--(0,-2,0)--(0,0,0)--cycle;
\draw[fill=blue!20,opacity=0.2](0,0,2)--(0,0,0)--(0,-2,0)--(0,-2,2)--cycle;
\draw[fill=blue!20,opacity=0.1](2,0,2)--(2,0,0)--(0,0,0)--(0,0,2)--cycle;
\draw[fill=blue!20,opacity=0.4](2,0,2)--(2,0,0)--(2,-2,0)--(2,-2,2)--cycle;
\draw[fill=blue!20,opacity=0.4](0,0,2)--(2,0,2)--(2,-2,2)--(0,-2,2)--cycle;
\end{tikzpicture}}
\caption{$I_{kin}$ represents dumbbell rotation.The blue cube is a cell of the lattice. The green cube is a cell of the dual lattice.}
\label{fig:dumb}
\end{figure}
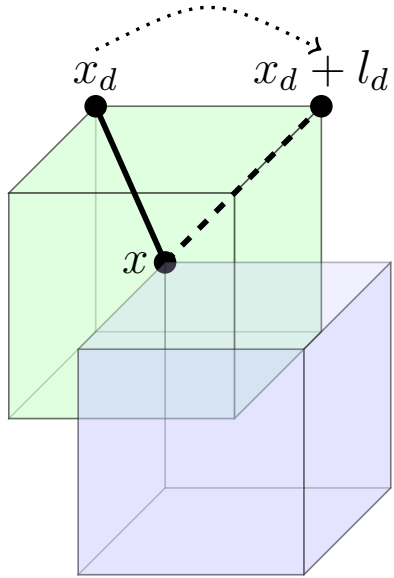

\section{Dumbbell Fields and Discrete Symmetries}

\ \ \ Kantor and Susskind introduced dumbbell fields $\psi (x,x_d)$ which depend on a point $x$ in the lattice and a point $x_d$ in the dual lattice. One such dumbbell is shown in figure \ref{fig:dumb}.  These dumbbells couple to the lattice gauge fields $a$ and the dual lattice gauge fields $b$. KS write two kinds of kinetic terms: translations, which move both the lattice point and the dual lattice point, and rotations, which move only one or the other.  We believe, since one can perform a translation by composing two rotations, that only one is necessary.  The analysis in this paper will take place far from the continuum limit, so we will not have an opportunity to test this conjecture.  We divide the action into three parts:the kinetic part which represents dumbbell rotation, the mass term,and the Chern-Simons term.

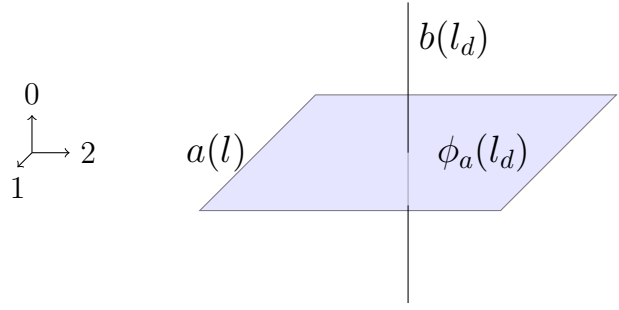
\begin{figure}[h]
\centering
\scalebox{1}{
\begin{tikzpicture}
\draw[fill=blue!20,opacity=0.5](2,0,2)--(2,0,-2)--(-2,0,-2)--(-2,0,2)--cycle;
\draw (0,-2,0)--(0,-0.7,0);
\draw [opacity=0.3](0,-0.7,0)--(0,0,0);
\draw (0,0,0)--(0,2,0);
\node at (0,1.5,0)[right]{\large $b(l_d)$};
\node at (1,0,0){\large $\phi_{a}(l_d)$};
\node at (-2,0,0)[left]{\large $a(l)$};

\draw[->] (-5,0,0)--(-4.5,0,0);\node at (-4.5,0,0)[right]{2};
\draw[->](-5,0,0)--(-5,0.5,0);\node at (-5,0.5,0)[above]{0};
\draw[->](-5,0,0)--(-5,0,0.5);\node at (-5,0,0.5)[below]{1};
\end{tikzpicture}}
\caption{Illustration of the Chern-Simons term. $a(l)$ lives on the lattice link $l$. $\phi_{a}(l_d)$ is the loop sum of $a(l)$ around the plaquette that $l_d$ penetrates. The coordinate system on the left shows the positive directions.}
\label{fig:cs}
\end{figure}

\beq I_{kin} =& \sum_{x,x_d,l} \bar{\psi} (x+l,x_d) e^{i Q_a a(l)} \psi (x, x_d) + \sum_{x,x_d,l_d} \bar{\psi} (x,x_d+l_d) e^{i Q_b b(l_d)} \psi (x, x_d)+h.c.\\
I_m=& m^2 \sum_{x,x_d} \bar{\psi} (x,x_d) \psi (x,x_d)\\
I_{CS}=&\sum_{l_d}b(l_d)\phi_{a}(l_d)
\label{eq:2} 
 \eeq

$x$ runs over all lattice points, $x_d$ runs over all dual lattice points on the dual cube surrounding $x$. $l$ is a lattice link, $l_d$ is the dual lattice link. We also use the notation $l_j, l_{dj} (j=0,1,2)$ to represent links in the $j$ direction. $I_{kin}$ represents dumbbell rotation(figure \ref{fig:dumb}). In $I_{kin}$, $l$ runs over the positive unit vectors such that the dual cube surrounding $x$ and the dual cube surrounding $x+l$ share the point $x_d$, $l_d$ runs over the positive unit vectors such that the cube surrounding $x_d$ and the cube surrounding $x_d+l_d$ share the point $x$. $a$ field lives on the lattice, and $a(l)$ is the $a$ field on lattice link $l$. $b$ field lives on the dual lattice, and $b(l_d)$ 
is the $b$ field on dual link $l_d$. $\phi_{a}(l_d)$ is the loop sum of $a(l)$ around the plaquette that $l_d$ penetrates. $Q_a,Q_b$ are the fictitious charges carried by two ends of the dumbbell, and we assume that $Q_a$ only couples to $a$, $Q_b$ only couples to $b$. $m$ is the dumbbell mass. Note that there is no Chern-Simons level in $I_{CS}$, because it can be absorbed into $Q_b$.   That is, the periodicity of $b$ is really determined by the quantization rule for the charges, $Q_b$, and we can absorb an integer multiple of the CS action by rescaling $b$ and making the appropriate change of $Q_b$.  

Under time-reversal operation T, $i\rightarrow -i$ and the spacial components of the gauge field reverses sign, $a(l_1),a(l_2)\rightarrow -a(l_1),-a(l_2)$ and $b(l_{d1}),b(l_{d2})\rightarrow -b(l_{d1}),-b(l_{d2})$. The dumbbell field is unchanged. Putting these into equation \ref{eq:2}, we find that the terms in the dumbbell action $I_{kin}+I_m$ that transform are 
\beq
&\bar{\psi}(x+l_0,x_d)e^{iQ_aa(l_0)}\psi(x,x_d)+\bar{\psi}(x,x_d)e^{-iQ_aa(l_0)}\psi(x+l_0,x_d)\\&+\bar{\psi}(x,x_d+l_{d0})e^{iQ_bb(l_{d0})}\psi(x,x_d)+\bar{\psi}(x,x_d)e^{-iQ_bb(l_{d0})}\psi(x,x_d+l_{d0})\label{eq:3}
\eeq
 It becomes
 \beq
&\bar{\psi}(x'-l_0,x'_d)e^{-iQ_aa(l_0)}\psi(x',x'_d)+\bar{\psi}(x',x'_d)e^{iQ_aa(l_0)}\psi(x'-l_0,x'_d)\\&+\bar{\psi}(x',x'_d-l_{d0})e^{-iQ_bb(l_{d0})}\psi(x',x'_d)+\bar{\psi}(x',x'_d)e^{iQ_bb(l_{d0})}\psi(x',x'_d-l_{d0})
\label{eq:4}\eeq
 where $x'=(-x_0,x_1,x_2), x'_d=(-x_{d0},x_{d1},x_{d2})$. Eq. \ref{eq:4} reduces to eq. \ref{eq:3} if we use another set of variables $x''=x'-l_0, x''_d=x'_d-l_{d0}$ to label the lattice sites. The dumbbell term is T-even.

Under parity operation P, we reverse one spacial direction. Choose the 1 direction, for example. One spacial component of the gauge fields reverse sign, $a(l_1)\rightarrow -a(l_1), b(l_1)\rightarrow -b(l_1)$. The terms in $I_{kin}+I_m$ that transform are 
\beq
&\bar{\psi}(x+l_1,x_d)e^{iQ_aa(l_1)}\psi(x,x_d)+\bar{\psi}(x,x_d)e^{-iQ_aa(l_1)}\psi(x+l_1,x_d)\\&+\bar{\psi}(x,x_d+l_{d1})e^{iQ_bb(l_{d1})}\psi(x,x_d)+\bar{\psi}(x,x_d)e^{-iQ_bb(l_{d1})}\psi(x,x_d+l_{d1})\label{eq:5}
\eeq
which becomes
\beq
&\bar{\psi}(x'-l_1,x'_d)e^{-iQ_aa(l_1)}\psi(x',x'_d)+\bar{\psi}(x',x'_d)e^{iQ_aa(l_1)}\psi(x'-l_1,x'_d)\\&+\bar{\psi}(x',x'_d-l_{d1})e^{-iQ_bb(l_{d1})}\psi(x',x'_d)+\bar{\psi}(x',x'_d)e^{iQ_bb(l_{d1})}\psi(x',x'_d-l_{d1})\label{eq:6}
\eeq
where $x'=(x_0,-x_1,x_2), x'_d=(x_{d0},-x_{d1},x_{d2})$.  Eq. \ref{eq:6} reduces to eq. \ref{eq:5} if we use another set of variables $x''=x'-l_1, x''_d=x'_d-l_{d1}$ to label the lattice sites. The dumbbell term is P-even.

Under charge-conjugate operation C, $\psi\rightarrow \bar{\psi}, \bar{\psi}\rightarrow (-1)^f\psi$, where $f=1$ for a fermionic dumbbell field, $f=0$ for a bosonic dumbbell field. $a\rightarrow -a, b\rightarrow -b$.The transformation of $I_m$ is 
\beq
m^2\bar{\psi}(x,x_d)\psi(x,x_d)\overset{C}{\longrightarrow} (-1)^f m^2\psi(x,x_d)\bar{\psi}(x,x_d)=m^2\bar{\psi}(x,x_d)\psi(x,x_d)
\eeq
The transformation of $I_{kin}$ is 
\beq
\ \ \ &\bar{\psi} (x+l,x_d) e^{i Q_a a(l)} \psi (x, x_d) + \bar{\psi} (x,x_d+l_d) e^{i Q_b b(l_d)} \psi (x, x_d)\\&
+ \bar{\psi} (x,x_d) e^{-i Q_a a(l)} \psi (x+l, x_d) + \bar{\psi} (x,x_d) e^{-i Q_b b(l_d)} \psi (x, x_d+l_d)\\
\overset{C}{\longrightarrow}&
 (-1)^f\psi(x+l,x_d) e^{-i Q_a a(l)} \bar{\psi} (x, x_d) + (-1)^f\psi (x,x_d+l_d) e^{-i Q_b b(l_d)} \bar{\psi} (x, x_d)\\&
+(-1)^f\psi (x,x_d) e^{i Q_a a(l)} \bar{\psi} (x+l, x_d) + (-1)^f\psi (x,x_d) e^{i Q_b b(l_d)}\bar{\psi} (x, x_d+l_d)
\\
 =\ \  &\bar{\psi} (x,x_d) e^{-i Q_a a(l)} \psi (x+l, x_d) + \bar{\psi} (x,x_d) e^{-i Q_b b(l_d)} \psi (x, x_d+l_d)
\\&+\bar{\psi} (x+l,x_d) e^{i Q_a a(l)} \psi (x, x_d) + \bar{\psi} (x,x_d+l_d) e^{i Q_b b(l_d)} \psi (x, x_d)
\eeq
which also returns to its original form. Both $I_{kin}$ and $I_m$ are C-even.

The Chern-Simons term is
\beq
I_{CS}=&\sum_{l_d}b(l_d)\phi_a(l_d)\\=&\sum_{l_{d0}}b(l_{d0})(\Delta_1a(l_2)-\Delta_2a(l_1))
+\sum_{l_{d1}}b(l_{d1})(\Delta_2a(l_0)-\Delta_0a(l_2))\\&
+\sum_{l_{d2}}b(l_{d2})(\Delta_0a(l_1)-\Delta_1a(l_0))\label{eq:9}
\eeq
where $\Delta$ is the lattice difference. Under C operation, both $a(l)$ and $b(l_d)$ reverse sign, so $I_{CS}$ doesn't change. If we reverse the spacial direction 1, $\Delta_1\rightarrow -\Delta_1, a(l_1)\rightarrow -a(l_1), b(l_{d1})\rightarrow -b(l_{d1})$, and it is obvious from equation \ref{eq:9} that $I_{CS}$ is odd under P. The T operation sends $\Delta_0\rightarrow -\Delta_0, a(l_1)\rightarrow -a(l_1), a(l_2)\rightarrow -a(l_2), b(l_{d1})\rightarrow -b(l_{d1}), b(l_{d2})\rightarrow -b(l_{d2})$. Put these into equation \ref{eq:9}, we find $I_{CS}$ is also odd under T. The violation of P and T tells us that the Chern-Simons term cannot appear in systems that preserve these symmetries. These terms are allowed in quantum Hall system, because the appearance of external magnetic field already breaks P and T. We have followed the condensed matter convention to define time reversal operation \cite{tong}, where $a(l_0),b(l_0)$ are even while the spacial components are odd.

\section{Wilson Loops and Jain's fractions}
\subsection{Wilson Loops}
\begin{figure}[h]
\centering
\scalebox{1}{
\begin{tikzpicture}
\draw[opacity=0.5](-0.6,0,-2)--(2,0,-2)--(2,0,0);
\draw[opacity=1](-0.6,0,-2)--(-2,0,-2)--(-2,0,2)--(2,0,2)--(2,0,0);
\draw[opacity=1](0.6,0,2)--(2,0,2)--(2,0,0);
\draw[fill=blue!20,opacity=0.5](2,0,-2)--(2,0,2)--(-2,0,2)--(-2,0,-2)--(2,0,-2);
\draw[fill=green!20,opacity=0.5](0,-0.7,0)--(0,-2,0)--(4,-2,0)--(4,2,0)--(0,2,0)--(0,0,0);
\draw (0,2,0)--(4,2,0)--(4,-2,0)--(0,-2,0)--(0,-0.7,0);
\draw [opacity=0.3](0,-0.7,0)--(0,0,0);
\draw (0,0,0)--(0,2,0);
\node at (2,0,0){\large $-1$};
\node at (0,0,2)[below]{\large $+1$};
\node at (-2,0,0)[left]{\large $+1$};
\node at (0.7,0,-2)[above]{\large $-1$};
\node at (0,0,0){\large $+1$};
\node at (2,2,0)[above]{\large $+1$};
\node at (4,0,0)[right]{\large $-1$};
\node at (2,-2,0)[below]{\large $-1$};
\draw [line width=0.5mm,->](-2,0,-0.1)--(-2,0,0);
\draw [line width=0.5mm,->](4,0,0)--(4,-0.1,0);

\draw[->] (-5,0,0)--(-4.5,0,0);\node at (-4.5,0,0)[right]{2};
\draw[->](-5,0,0)--(-5,0.5,0);\node at (-5,0.5,0)[above]{0};
\draw[->](-5,0,0)--(-5,0,0.5);\node at (-5,0,0.5)[below]{1};
\end{tikzpicture}}
\caption{An example of the $J$ configuration. Numbers around the blue square are $J_a(l)$, while numbers around the green square are $J_b(l_d)$. The coordinate system on the left shows the positive directions. The configuration represents one $a$ loop and one $b$ loop penetrating each other.}
\label{fig:J}
\end{figure}
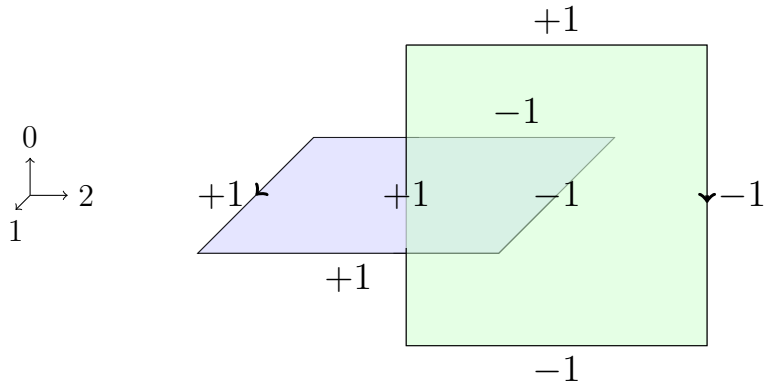

\ \ \ The partition function is 
\beq
Z=\int \dd a\int \dd b\int \dd\bar{\psi}\int \dd\psi \ e^{i(I_{kin}+I_m+I_{CS})}
\eeq
Integrating over the dumbbell fields we get a determinant of an operator $(m + D)$, where $D$ is a sum of gauge invariant shift operators on the lattice and dual lattice. 
\beq
Z=\int \dd a\int \dd b\ det(m+D)e^{iI_{CS}}
\eeq
If we expand the determinant in powers of $m^{-1}$ when $m$ is large, we obtain an expression for it as a sum over Wilson loops of the $a$ and $b$ fields. The detailed analysis can be found in section IV of reference\cite{bingnan}.  This expansion is obviously convergent on a finite lattice.  On a torus of equal lengths $L$ it contains terms of order $m^{-L} $, which are sensitive to the topology. Now we have reduced $Z$ into the form
\beq
Z=\sum_{\{J\}}Z[J]=\sum_{\{J\}}\int \dd a\int \dd b\ e^{i[Q_a\sum_l J_a(l)a(l)+Q_b\sum_{l_d}J_b(l_d)b(l_d)+\sum_{l_d}b(l_d)\phi_a(l_d)]} 
\label{eq:12}
\eeq
where $J_a,J_b\in \mathbb{Z}$ label the Wilson/t'Hooft loops. One example of the $J$ configuration is shown in figure \ref{fig:J}

As we will show below, integrals over the gauge potentials, $a,b$, of finite products of Wilson loops involve only linking numbers between $a$ loops and $b$ loops.  Thus the partition function and correlation functions of localized products of dumbbell operators have a convergent power series expansion in powers of $m^{-1}$.  The model is defined by analytic continuation in $m$ beyond the radius of convergence of this series.  

Now we will argue that for the $U(1)$ theory, certain charge pairs $(Q_a, Q_b)$ are equivalent to each other to all orders in the $1/m$ expansion.  Convergence of the expansion implies that the equivalence is true for a range of complex $m$ and is, at the very least, strong evidence that it is an exact property of the model.  Among these equivalences are those used in\cite{jain}\cite{HLR}\cite{fradkin} to construct the composite fermion mean field theory of the FQHE.  

If we introduce external magnetic field $A(l)$ and attach a real physical charge $Q$ to one end of the dumbbell, the action in equation \ref{eq:12} becomes
\beq
S[J]:=\sum_{l_d} [b(l_d)\phi_{a}(l_d)+Q_bJ_{b}(l_d)b(l_d)+QJ_{b}(l_d)A(l_d)]+\sum_{l}Q_aJ_{a}(l)a(l)
\eeq
 First we assume both $a$ and $b$ are $\mathbb{R}$ variables and calculate $Z[J]$ defined in equation \ref{eq:12}. Integrating over $b(l_d)$ we get a $\delta$ function for $\phi_a(l_d)$
\beq
Z[J]=&Z[0]^{-1}\int \dd a\int \dd b e^{iS[J]}\\=&Z[0]^{-1}\int \dd a\  \delta(\phi_a(l_d)+Q_bJ_b(l_d))e^{i\sum_{l}Q_aJ_a(l)a(l)+i\sum_{l_d}QJ_{b}(l_d)A(l_d)
}\label{eq:14}
\eeq
Using the lattice Stokes' theorem, we can replace the loop sum $\sum_l J_a(l)a(l)$ with a surface sum $\sum_{\l_d}\bar{J}_{a}(l_{d})\phi_{a}(l_d)$, where $\bar{J}_{a}(l_d)=1$ when $l_d$ penetrates a plaquette inside the $J_a(l)$ loop. The integral $\int \dd a$ can be replaced with $\int \dd\Lambda_a\ \dd\phi_a$, where $\Lambda_a$ is a pure gauge and $\phi_a$ bounds no net charge. The integral $\int \dd\Lambda_a$ is canceled by $Z[0]^{-1}$, so we have
\beq
Z[J]&=Z[0]^{-1}\int \dd\Lambda_a\ \dd\phi_a\ \delta(\phi_a(l_d)+Q_bJ_b(l_d))
e^{i\sum_{\l_d}Q_a\bar{J}_{a}(l_{d})\phi_{a}(l_d)+i\sum_{l_d}QJ_{b}(l_d)A(l_d)}
\\&=e^{-iQ_aQ_b\sum_{l_d}\bar{J}_{b}(l_d)J_{a}(l_d)+i\sum_{l_d}QJ_{b}(l_d)A(l_d)
}
\eeq 
The sum $\sum_{l_d}\bar{J}(l_d)J(l_d)$ counts how many times $J_a$ loops penetrate $J_b$ loops, which is an integer. When the fictitious charge $Q_aQ_b=2n\pi$ where $n$ is an arbitrary integer, we have
\beq
Z[J]=e^{i\sum_{l_d}QJ_{b}(l_d)A(l_d)
}
\label{eq:zj}
\eeq
adding $a$ and $b$ fields to the action is equivalent to not adding them.\par

\par When $b\in U(1)$ while $a\in \mathbb{Z}$, integrating $\int_0^{2\pi}\dd b$ in equation \ref{eq:14} yields a Kronecker delta $\delta_{\phi_{a}(l_d)+Q_bJ_{b}(l_d)}$. Summing $a$ over $\mathbb{Z}$ again gives the result equation (\ref{eq:zj}). When $Q_aQ_b=2n\pi$, the system is again equivalent to the one without $a$ and $b$ fields. The only thing different is that $Q_a$ has to be an integer so that the action is invariant under the shift $a\rightarrow a+2\pi$. $Q_b$ is an angle. 

\subsection{Large Gauge Transformations}
\ \ \ When $a\in\mathbb{R}, b\in\mathbb{R}$, the invariance of $S_{CS}$ under large gauge transformation is straightforward to check. When $a\in U(1),b\in\mathbb{Z}$, the situation is a little subtle. In this case $Q_a\in\mathbb{Z}, Q_b\in U(1)$  in order to preserve the invariance of the action under a $2\pi$ shift. Suppose the lattice is on $\mathbb{T}^3$ and we do a large gauge transformation along the 0 direction \beq b(l_{d0})\rightarrow b(l_{d0})+\Delta_0\omega(x_{d0})\label{eq:a1}\eeq where $\omega(x_{d0})\in\mathbb{Z}$ is an integer function on the dual lattice. The dumbbell wavefunction obtains an additional phase factor $e^{iQ_b\omega(x_{d0})}$. In order to make the phase factor single-valued, we should have \beq Q_b\omega(x_{d0}+N_0)-Q_b\omega(x_{d0})\in 2\pi\mathbb{Z}\label{eq:a2}\eeq
where $N_0$ is the number of lattice sites along the 0 direction. When $Q_b=\frac{p}{q}2\pi$ is a fraction of $2\pi$, the large gauge transformation should satisfy $\omega(x_{d0}+N_0)-\omega(x_{d0})=qs\in q\mathbb{Z}$. The variation of $S_{CS}$ is 
\beq
\delta S_{CS}=\sum_{l_0}\Delta_0\omega(x_{d0})\phi_a(l_0)=\sum_{x_{d0}}\Delta_0\omega(x_{d0})\sum_{x_{d1},x_{d2}}\phi_a(l_0)
\eeq
where $x_{d1},x_{d2}$ run over a closed spacial surface. If the spacial surface encloses a $Q_b$ charge then $\sum_{x_{d1},x_{d2}}\phi_a(l_0)=-Q_b$, so 
\beq
\delta S_{CS}=-Q_b\sum_{x_{d0}}\Delta_0\omega(x_{d0})=-Q_b[\omega(x_{d0}+N_0)-\omega(x_{d0})]=-2\pi ps\in 2\pi\mathbb{Z}
\eeq
and $e^{iS_{CS}}$ is unchanged.

When $Q_b$ is not a fraction of $2\pi$, equation \ref{eq:a2} can never be satisfied and there is no large gauge transformation for $b$. Fortunately, in this paper we want $Q_aQ_b=2n\pi$ so $Q_b=\frac{2n\pi}{Q_a}$ always satisfies the requirement.

\subsection{Jain's Fractions}
\ \ \ When we go to Hamiltonian formalism, we first integrate out the temporal components  $a(l_0)$ and $b(l_{d0})$. This results in two constraints 
\beq
\phi_{b}(l_0)&=-Q_aJ_{a}(l_0)\\
\phi_{a}(l_{d0})&=-Q_bJ_{b}(l_{d0})
\eeq
In the continuous limit, the dumbbell becomes a point particle, and the two constraints become
\beq
F_b&=-Q_a\rho\\
F_a&=-Q_b\rho
\eeq
where $F_a$ is the field strength of $a$, $F_b$ is the field strength of $b$, $\rho$ is the particle density. The constraints tell us that every particle carries $b$ flux $-Q_a$ and $a$ flux $-Q_b$. The particle carries both $Q_a$ charge and $Q_b$ charge. As we switch two particles, the phase change caused by $a$ and $b$ is \beq \theta=(-Q_aQ_b-Q_bQ_a)/2=-Q_aQ_b\eeq \\When $Q_aQ_b=2n\pi$, the statistics doesn't change.
\par The effective filling fraction for a uniform density profile is
\beq
\nu_{eff}=&\frac{2\pi\rho}{QF_A+Q_aF_a+Q_bF_b}=\frac{2\pi\rho}{QF_A-Q_aQ_b\rho-Q_bQ_a\rho}\\=&\frac{2\pi\rho}{QF_A-4n\pi\rho}=\frac{\nu}{1-2n\nu}
\eeq
where $F_A$ is the external field strength, $\nu=\frac{2\pi\rho}{QF_A}$ is the bare filling fraction, and $Q_aQ_b=2n\pi$ has been assumed. This agrees with the Jain's fractions derived in\cite{jain}\cite{HLR}\cite{fradkin}.

We've argued that the unambiguous agreement of all dumbbell loop expectations values for two different values of charges implies that the models are actually equivalent non-perturbatively.  Previous authors have argued that the mean field theory where we treat interactions perturbatively when the charges $Q_a, Q_b$ match the filling fraction as above,  gives a good approximation to the exact dynamics of the model. In principle our formalism could lead to a more systematic and rigorous version of those arguments, but we leave that task for future work.

\section{Higher Forms and Higher Dimensions}
\ \ \ 
Let $G$ be an abelian group and $G_D$ the dual group of its irreducible representations.   In $d$ dimensions we can defined generalized CS actions of the form 
\beq I = i \sum_P b_{p} (P_D) da_{d - p - 1} (P)  \eeq  Here $P$ is a $d-p$ dimensional face of a $d$ dimensional hypercube and $P_D$ is the $p$ dimensional face of the dual hypercube, which penetrates $P$.  $b_p$ belongs to the additive group of $G_D$ while $a_{d-p-1}$ is in the additive group $G$.  Models of this form were first studied by\cite{kapustin}, for the case $G_D = G$.  

We can consider dumbbell fields $\psi (F_{d-p-2} , F^D_{p-1}) $, which live on minimal size faces of the lattice and dual lattice, with the indicated dimensions.  Note that each such face belongs to two different hypercubes so we can add terms to the action that allow the dumbells to move.  Note however that these terms cannot be bilinear in the field, since each $p$ face has more than a single pair of $p - 1$ "edges" if $p \geq 2$.  At a minimum, we must have a term involving $\psi (F_{d-p-2} , F^D_{p-1})$ on half of the face pairs, and a field $\bar{\psi}$ with opposite gauge transformation properties, on the other half.

As an example, consider the $ \int b_2 d a_1$ action in $4$ dimensions.  The dumbbell fields live on points of the lattice and links of the dual lattice.  We assign $\psi$ and $\bar{\psi} $  to all dual links.  The kinetic term in the action is then

\beq I_{kin} =& \sum_x \sum_{F_2^{D} (x)} e^{i b(F_2^D)} \prod_{F_1^D \in F_2^D} \bar{\psi} (x, F_1^D) \psi (x, \tilde{F}_1^D)   + \sum_{x,L} \bar{\psi} (x, F_1^D  (x,L)) e^{i a (L)} \psi (x + L, F_1^D (x,L) )\\&+h.c.\eeq  $L$ runs over all positive unit vectors such that the hypercube surrounding $x$ and the hypercube surrounding $x+L$ share a link denoted by $F_1^D(x,x+L)$. Note that in this case, if both signs of $b_2$ charge exist on all links, we can have a bilinear kinetic term.  If we assign $\psi$ fields only to positively oriented links and $\bar{\psi}$ fields to negatively oriented links the bilinear kinetic term is not allowed.   It is also forbidden if both $p$ and $d - p -1$ are larger than $1$.

There's an interesting analogy between models with the minimal number of $\psi$ and $\bar{\psi}$ fields, and chiral fermions carrying gauge charge\cite{bzcage}.  If the model does not have a correlation between gauge charge and orientation of the $p - 1$ and $d - p -2$ faces, then we can write a gauge invariant  "mass" term
\beq I_m = m \sum_{x, F_{p-1}^D (x) , F_{d - p - 2} (x) } \bar{\psi} (x, F_{p-1}^D (x) , F_{d - p - 2} (x) \psi (x, F_{p-1}^D (x) , F_{d - p - 2} (x) ) . \eeq   Treating the matter fields as either bosons or fermions, we take the large $m$ limit and obtain an expansion of the model in inverse powers of $m$.  Each power can be associated with sums over configurations of $p - 1, d - p - 2$ brane world volumes.  The gauge field integrals can then be carried out, giving us topological intersection numbers of these world volumes.  All of this can be done for an arbitrary abelian group, with dumbbell fields transforming in pairs of group and dual group representations.  When the fields are fermions, the expansion is convergent.  

\section{Conclusions}
\ \ \ 
We have extended the dumbbell formalism of Kantor and Susskind to construct lattice regularizations of CS theories for arbitrary abelian groups, and used it to provide a rigorous foundation for composite fermion theories of the FQHE.  The key new feature was the recognition that the gauge group on the dual lattice had to be the dual group to the of the original CS gauge potential.  We also proposed a generalization of the dumbbell formalism to higher form gauge theories in arbitrary dimensions.  This also revealed a notion of chiral matter fields for these higher form gauge fields, whose implications remain to be explored.

Finally, we note that it might be possible to extend our approach to non-abelian CS models.  In the absence of sources for $b$, the CS action sets the field strength of $a$ equal to zero on each plaquette of the lattice.  The identity

\beq
\frac{1}{|G|}\sum_R \chi_R(U(P))d_R=\delta(U(P))
\eeq 
for a compact group, where $d_R$ is the dimension of the representation,
$R$ runs over all the irreducible unitary representations, $\chi_R (U) = {\rm Tr}_R (U) $ and the 
$\delta$ function on the group sets $U_P = 1$, is the generalization of that flatness condition to all compact groups.  This suggests that the appropriate generalization of CS theory has a variable taking values in the irreducible representations of the group on each dual lattice link and that 
\beq e^{i I_{CS}} = \prod_P \sum_R U(P_D) \chi_R (U(P))\frac{d_R}{|G|} . \eeq
The question then arises of whether there is an analog of the $b$ Wilson loops, when $e^{ib}$ is a representation of a non-abelian group. Given a group element $g$ and a choice of the $R(P_d)$ for every dual link, we can define the product of the characters $\chi_{R(P_D)} (g)$ around the loop.  This can be evaluated using the Clebsch-Gordan expansion of the product of two characters.  We will leave the investigation of this proposal to future work.

\section*{Acknowledgment}
\ \ \ We thank N.Seiberg, G.Moore, S.Shao, S.Kivelson, E.Fradkin for reading the draft. This work  is partially supported by the Department of Energy under grant DOE
SC0010008.

\end{document}